# Mechanical characterization and failure modes in the peeling of adhesively bonded strips from a plastic substrate

Hamed Zarei[1*], Maria Rosaria Marulli [1], Marco Paggi[1], Riccardo Pietrogrande[2], Christoph Üffing[3], and Philipp Weißgraeber[3]

[1]IMT School for Advanced Studies Lucca, Piazza San Francesco 19, 55100 Lucca, Italy

[2]Automotive Electronics, Engineering Technology Polymers (AE/ETP), Robert Bosch GmbH, Renningen, Germany

[3]Corporate Research and Advance Engineering (CR), Robert Bosch GmbH, Renningen, Germany

**Abstract**

With the aim of understanding failure modes in the peeling of silicone-based adhesive joints and, in particular, the occurrence of adhesive or cohesive failure, an experimental campaign has been conducted by considering plastic substrates with different surface roughness. A flexible strip has been bonded onto such substrates using a silicone adhesive, by controlling its thickness. Peeling tests with 90º and 180º peeling angle configurations have been performed and the effect of joint parameters, such as surface roughness and adhesive thickness, onto the adhesion energy and the failure mode are herein discussed in detail. Experimental results show that the failure mode varies in each peeling test configuration such that in the case of 180º peeling test there is mainly cohesive failure, while for 90º peeling angle, a combination of adhesive and cohesive failure occurs. Moreover, due to the presence of different failure modes in each peeling configuration, the substrate roughness can increase the adhesion energy only in 90º peeling tests.

---

[1]*Corresponding author. E-mail: hamed.zarei@imtlucca.it

# 1. Introduction

Adhesion is the state when two dissimilar materials are in contact and in which mechanical force or work can be transferred across their interface. The joining technique could be based on a physical or a chemical bond, depending on the adhesive and the adherend. Adhesives are being used extensively in various industries due to their noticeable performance. Moreover, fracture characterization of adhesives received a great deal of attention by researchers [1–3] and enhancement of interfacial properties is one of the key research topics in this community [4–6]. The peel test is one of the most frequently used method to assess the interfacial mechanical properties between adhesive and adherend, and the peel strength is a measure of the force required to separate the peel arm from a substrate [7].

There are several parameters affecting the peeling response of adhesive joints, such as intrinsic adhesive properties including strength and viscoelasticity [8–10], surface roughness [11,12] and environmental factors [13–15]. Wei and Hutchinson [8] determined the correlation between the peel strength and the work of interface adhesion through a Cohesive Zone Model (CZM) for fracture. Kim et al. [9] studied the mechanical properties effects in the 90° peeling test. Accordingly, they found that, to evaluate the peel strength, the elastic-plastic deformation of the adhesive layer should be taken into account. The adhesion mechanism of a viscoelastic thin-film on a rigid substrate was investigated in [10] through a combined experimental-analytical approach. Peng et al. [10] concluded that the energy release rate of the film/substrate system increases nonlinearly while increasing the loading rate. Peng and Chen [11], and Huber et al. [12] studied the influence of surface roughness on the gecko adhesion for bio-inspired application. They demonstrated that the adhesion force would decrease or increase depending on the surface roughness level. Datla et al. [13–15] assessed the environmental factors that can negatively affect the bonded joint over time, such as temperature, and moisture. Enhancing the fatigue threshold while increasing the crack growth rate in a raised temperature environment under dry conditions can be mentioned as one of the major results.

Essentially, in the scientific community, in terms of surface topography investigation, the adhesive problems can be classified into two categories: (i)- weak adhesion or van der Waals adhesion [16,17], and (ii)- strong adhesion or adhesion including intermediate layer [18]. Cho et al. [18] studied the Aluminum adherend roughness effect on the strength of RTV88 adhesive strength through tensile-shear tests. They introduced effective area, peel failure area, and cohesive failure area to explain the surface roughness effect on adhesive strength. The focus of the current article is the differentiation of the failure modes of silicone-based adhesive joints through 90° and 180° peeling tests, examining the effect of adhesive thickness and substrate roughness. We focus in particular on two different failure modes: Adhesive Failure (AF), referring to an interfacial bond failure mode, which may occur at the interface between the adhesive and the adherends (the substrate or the strip); Cohesive Failure (CF), occurring when a crack travels within the finite thickness of the adhesive and, at failure, thin layers of the adhesive remain attached onto both adherends. To the best of the authors' knowledge, these aspects of silicone-based adhesive joints have not been hitherto investigated. To this end, Section 2 describes the experimental program including sample preparation and peeling test setups. Section 3 contains discussions on experimental observations, the post-mortem analysis of peeled off specimens, as well as substrate roughness contribution to the peeling response in detail. Eventually, the main conclusions are provided in Section 4.

## 2. Materials and Methods

### 2.1. Specimen preparation

A black glass fibre reinforced Polybutylene Terephthalate (PBT-GF30) polymer-based, and a flexible and inextensible HELIOX PV FERON NEOX CPC 300 have been used as substrate and strip for the peeling tests, respectively. The substrate dimensions were 120 mm × 25 mm × 2 mm, and the strip dimensions were 5 mm wide × 0.3 mm thick. In addition, a black two-component silicone sealant cured at room temperature has been applied as adhesive. To assess the roughness effect, the substrates have been fabricated by inserting steel blocks in the last $25 \times 25$ mm$^2$ part with specified roughness into a mold form used in the injection molding process. Since the produced roughness might be different from the roughness of the steel block, roughness characterization of the polymeric surface has been carried out using the non-contact confocal profilometer LEICA DCM3D available in the MUSAM-Lab at the IMT School for Advanced Studies Lucca. In addition, a proper fixture has been designed within this research project to bond the strip onto the substrate, which was capable to control adhesive thickness during manufacturing. Fig. 1 shows the prepared sample with the nominal roughness $R_z$=7.5 μm and the adhesive thickness t=0.8 mm. The whole procedure of the sample preparation including substrate fabrication, surface roughness acquisition and strip bonding onto the substrate are addressed in detail in [19].

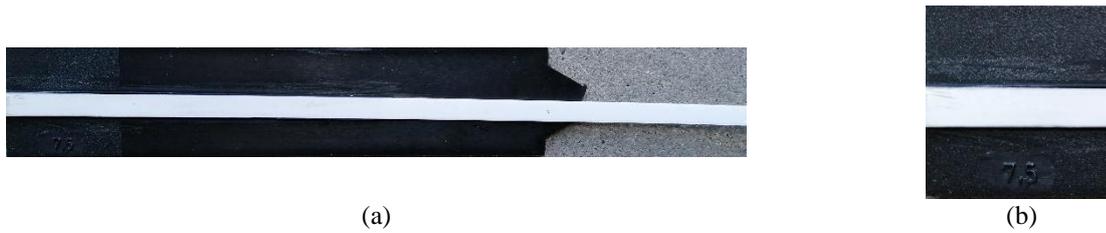

(a)                          (b)

*Fig. 1. The prepared sample for the peeling test. (a): the whole substrate with rough (left end) and smooth (on its right) portion, (b): detail of the rough part.*

### 2.2. Experimental procedures

As shown in Fig. 2, all the experiments have been carried out by employing the Zwick/Roell universal testing machine in two peeling test configurations, i.e. 90° and 180°, according to the ASTM- D1876 and C794-01 standards, respectively. It should be noted that, by properly adjusting the peeling set-up, a desired peeling angle can be achieved. The prepared specimen is placed between the designed clamp and base body of the peeling kit, while the free strip is fixed within the grip installed on the moving crosshead. The peeling force is acquired with a load cell mounted on the fixed crosshead while the peel extension is measured based on the absolute crosshead travel. The force-displacement data as well as test control parameters such as preload, loading rate, etc. are accessible through the testXpert II V 3.41 software interface. For the present study, both the 90° and 180° peeling responses of specimens with the adhesive thickness of t=0.8, 1.2, and 1.9 mm have been studied. Furthermore, five different substrates having the nominal roughness of $R_z$=0.7, 1.8, 3.0, 7.5 μm as well as a smooth substrate have been considered. For each test, at least three samples have been tested, for a total of 90 peeling test acquisitions.

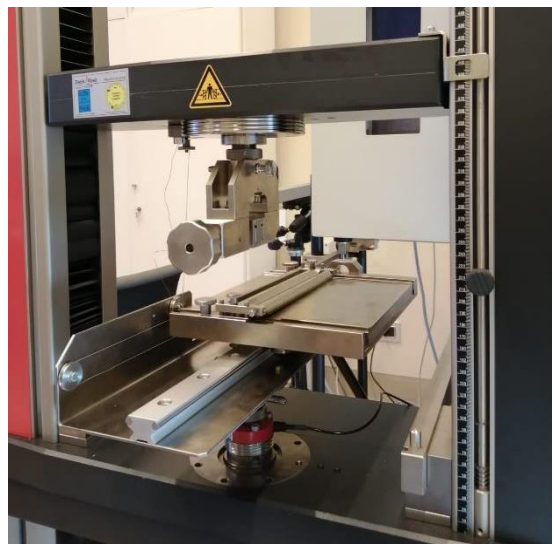

*Fig. 2. Peeling test setup.*

## 3. Results and Discussion

### 3.1. Experimental observations and post-mortem analysis

The fracture surface of the investigated specimens has been examined using the images of both strip and substrate after the peeling test. An image processing routine implemented in MATLAB has also been used to distinguish the damage pattern mode in either cohesive or adhesive failure (CF or AF) mode, and assess the percentage of each damage mode as well. Concretely, if the strip is white, it represents that there is an AF close to the strip. When the strip is black, the substrate surface should be analyzed to distinguish between CF or AF failure modes. The failure pattern modes have been depicted in Fig. 3, schematically.

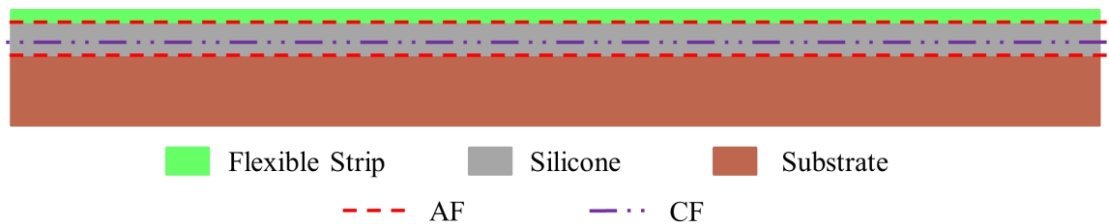

*Fig. 3. A schematic illustration of adhesive (AF) and cohesive (CF) failure modes.*

For instance, Figs. 4 and 5 show the 90° peeling response of three similar specimens (a, b, and c), and the post-peeled images of the strips and the substrates, respectively. All three specimens include a smooth substrate bonded by a 1.2 mm thick adhesive to the strip. In all such three cases, the damage pattern has initiated with an AF mode on the substrates and has made the first peak load followed by CF mode close to the strip. In case (a), after ending the initial AF stage, the crack has propagated and caused the second peak load. The peeling force has then reached the plateau state, which is related to the adhesion energy [20]. While in both (b) and (c) cases, once the plateau state was reached, AF on the substrate can be seen again in the middle of the specimen, which was responsible for the load drop. Finally, a mixed-failure mode, i.e. CF and AF, on the strip can be seen in case (b), leading to some drops in the peeling force level. On the other hand, almost the same peeling force level has been recorded due to the same failure pattern for the other two cases, (a) and (c).

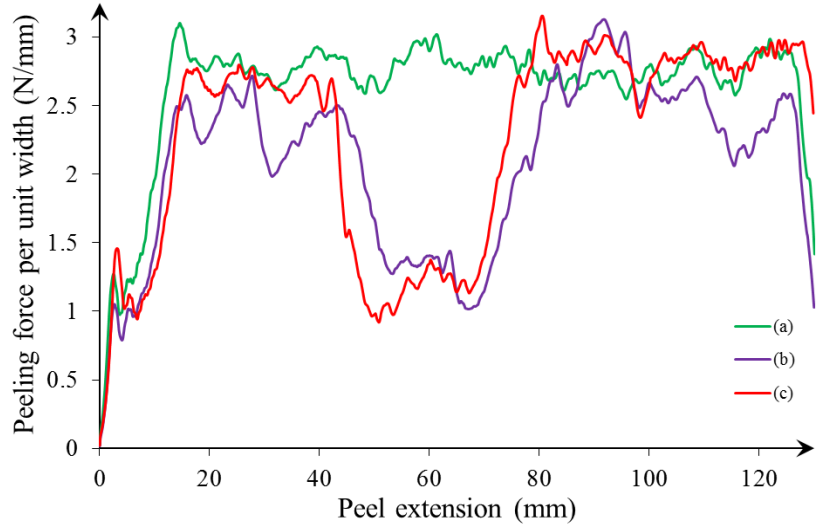

*Fig. 4. 90° peeling response of smooth specimens with the adhesive thickness t=1.2 mm.*

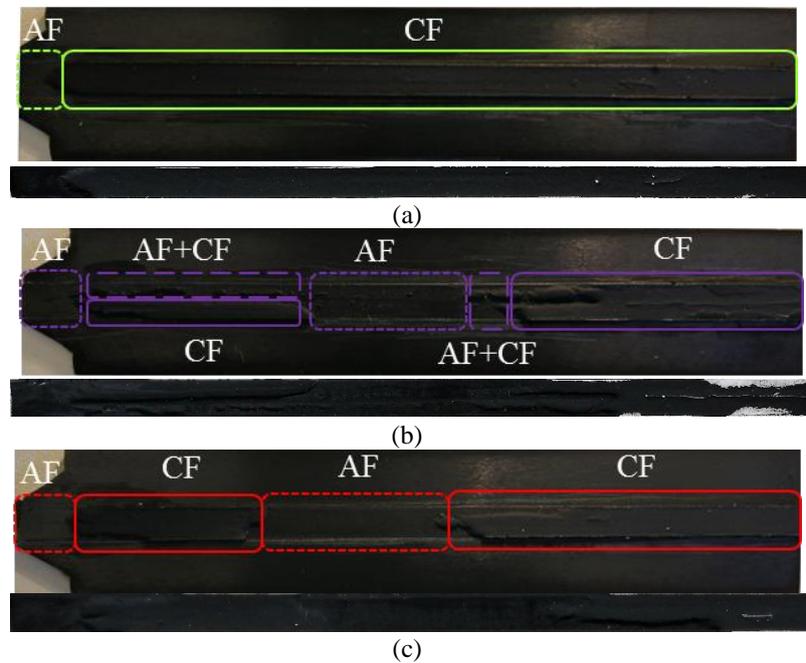

*Fig. 5. Post-peeled images of the smooth substrates and the strips through 90° peeling test with the adhesive thickness t=1.2 mm.*

As the next example, Fig. 6 illustrates the peeling force vs. displacement response of two peeled back specimens with the damage pattern produced on the strip. Accordingly, since the failure mode was almost the same for two cases, the force-displacement curves almost coincide in the initial part. Then, the changes in failure mode led to changes in the peeling force response, so that a sudden loading drop was observed when the AF mode on the strip occurs. Concretely, there are both AF and CF modes close to the strip in such two cases, and the corresponding CF ratio – the percentage of the black area on the strip which represents the CF- for the upper and lower strips shown in Fig. 6 are 80% and 65%, respectively.

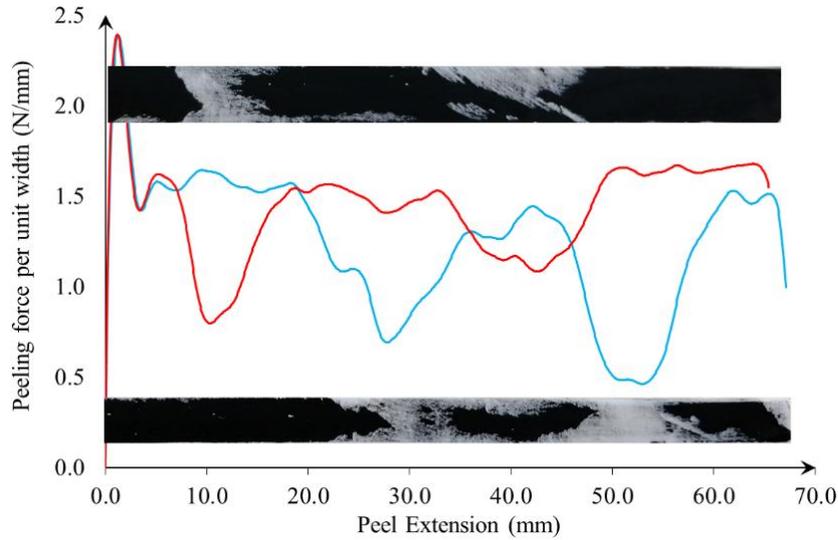

*Fig. 6. 180° peeling response of specimen with adhesive thickness t=1.9 mm and the damage pattern created on the strip.*

To assess the contribution of damage topography on the peeling force, the most common failure modes in each peeling test configuration i.e. 90° and 180°, have been considered and the corresponding results have been provided in Figs. 7 and 8.

In the first scenario, the 90° peeling response of specimens with the adhesive thickness t=0.8 mm has been evaluated. Experiments reveal that the mixed failure mode including CF close to the strip and AF on the substrate, which caused a ridge pattern on the middle of the post mortem sample surface (Figs. 7b and 7c), requires higher peeling forces compared to the perfect CF (Fig. 7a). The reason behind this phenomena could rely on the fact that in the case of mixed failure mode, there is larger surface creation as compared to the CF case in which a higher load is needed for the adhesive be fractured. In other words, more energy would be dissipated to propagate a crack in a mixed failure mode with a ridge pattern on the substrate.

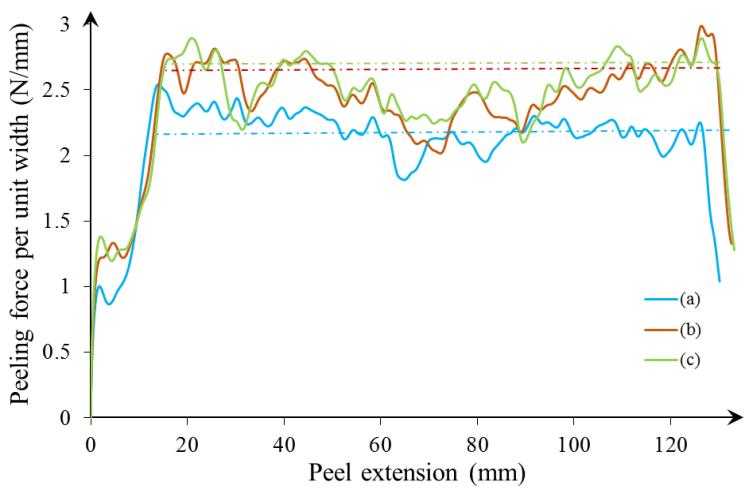
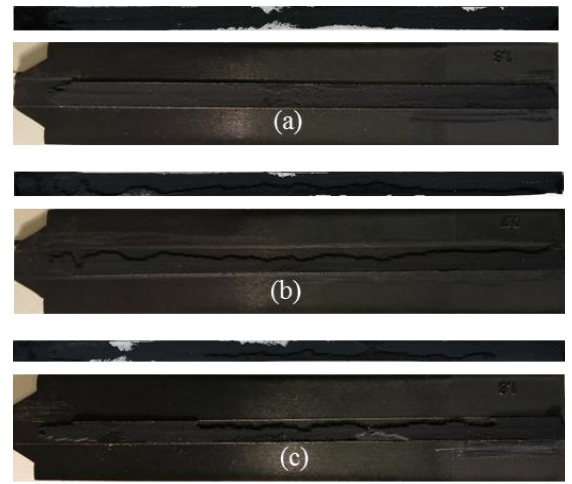

*Fig. 7. Damage topography contribution on the 90° peeling force of specimen with adhesive thickness t=0.8 mm: perfect CF vs. mixed failure mode.*

In the next scenario, CF and AF on the strip have been examined. Fig. 8 illustrates the peeling force vs. peel extension of the specimen with an adhesive thickness t=0.8 mm in the 180° peeling test configuration. In case (a), the damage pattern initiates with AF on the strip, followed by CF close to the strip, again AF, and ends with CF close to the strip. In case (b), there is almost CF close to the strip through the whole adhesive bond line, except a small AF portion in the last part of the interface. As is clear, CF requires more energy than AF for the present material combination, which has led to a higher peak load as compared to the adhesive failure at the beginning. However, the same plateau peeling force can be seen due to similar damage patterns, i.e. CF close to the substrate for both cases. It should be noted that the presence of the same AF on the substrate, in case (a), causes load drops such that the peeling force reaches the same level which is specified in Fig. 8.

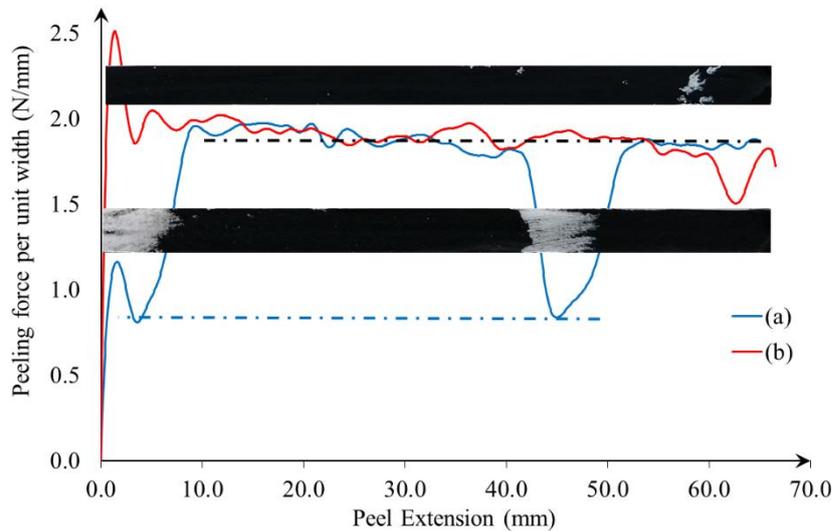

*Fig. 8. Damage topography contribution on the 180° peeling force of specimen with adhesive thickness t=0.8 mm: CF vs. AF on the substrate.*

### 3.2. Roughness effect

In this section, the effect of roughness on the peeling response of specimens through 90° and 180° peeling tests has been assessed. As discussed in detail in [19] for the 90° peeling test, the surface roughness can affect the adhesive strength if ($\alpha$) the ratio between the adhesive thickness and the root mean square of the roughness elevation meets the proposed criteria such that for those cases having $\alpha$ smaller than the critical value (i.e., $\alpha<246$), the surface roughness does matter to the adhesive peeling response. Otherwise, the substrate roughness effect would be negligible for this material combination. For example, Fig. 9 shows the peeling force vs. peel extension curve of specimens with different $R_z$ values and the adhesive thickness t= 0.8 mm for the 90° peeling test. Moreover, the corresponding α for the substrates with different $R_z$ values are collected in Table 1. It should be noted that for the samples including rough surface, the last 25 mm of peel extension corresponds to the rough part in the force- displacement curves. As is clear, for cases having α=199 and 70, corresponding to $R_z$=3.0 and 7.5 µm, there is a jump in the peeling force level in the rough part. In other words, substrate roughness leads to an enhancement in the adhesion energy. For the other cases, i.e. $R_z$=0.7, and 1.8 µm, the roughness effect is negligible.

To gain a deeper insight into the understanding of the failure modes behind the surface roughness effect, the same tests have been repeated for the 180° peeling test setup and the results are plotted in Fig. 10. It can be deduced that the surface roughness does not play a significant role on the adhesion strength in the 180° peeling test, while the different failure modes in the two cases play a substantial role. Once the strip is peeled off perpendicularly to the substrate surface, the crack propagates at the interface in the opening mode, while when peeling back the strip, mixed-mode fracture with dominant shear failure mode occurs. Therefore, based on the experimental results, only in Mode I, the surface roughness has a pronounced effect on the adhesion energy.

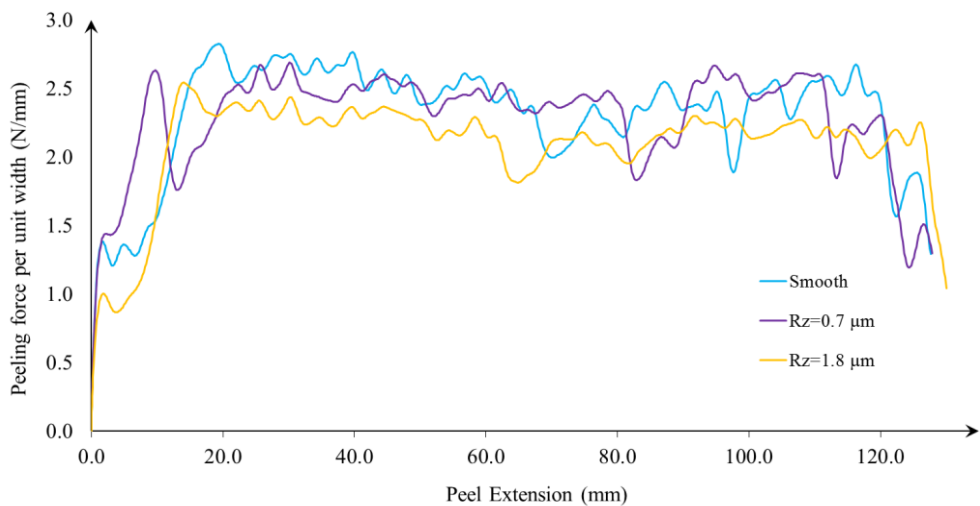

(a)

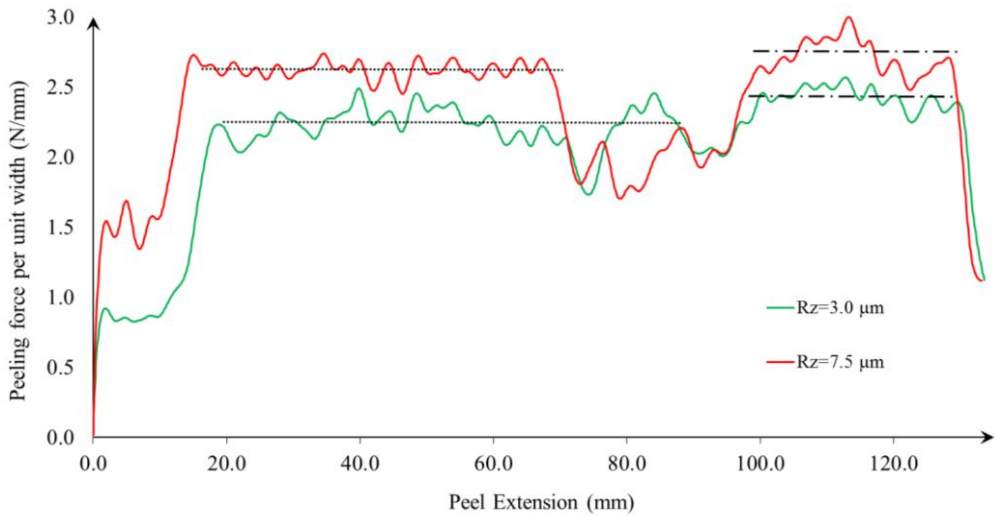

(b)

*Fig. 9. 90° peeling response of specimens with different $R_z$ values and adhesive thickness t=0.8 mm.*

*Table 1. Dimensionless adhesive thickness to substrate roughness ratio for the adhesive thickness t=0.8 mm.*

| $R_z$ (µm) | α |
|---|---|
| 0.7 | 363 |
| 1.8 | 246 |
| 3.0 | 199 |
| 7.5 | 70 |

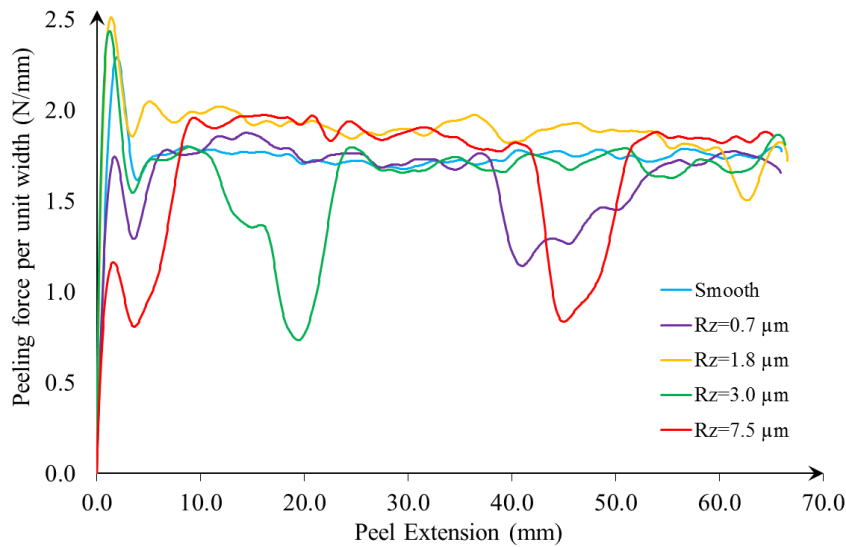

*Fig. 10. 180° peeling response of specimens with different $R_z$ values and adhesive thickness t=0.8 mm.*

The effect of surface roughness on the failure mode has been shown in Fig. 11. In the smooth part, damage initiates with almost a halved failure pattern in the lateral direction, including CF close to the strip and CF close to the substrate, followed by a transition to almost the same pattern along with the propagation of the crack. In other words, there is CF close to the strip at the top of the region, and then CF becomes close to the substrate later on, with a propagation through the adhesive thickness in the transition zone. Then, in the rough part, there is a mixed failure mode including both CF and AF close to the substrate, making a ridge pattern with tortuosity in the middle, ending with the CF close to the strip.

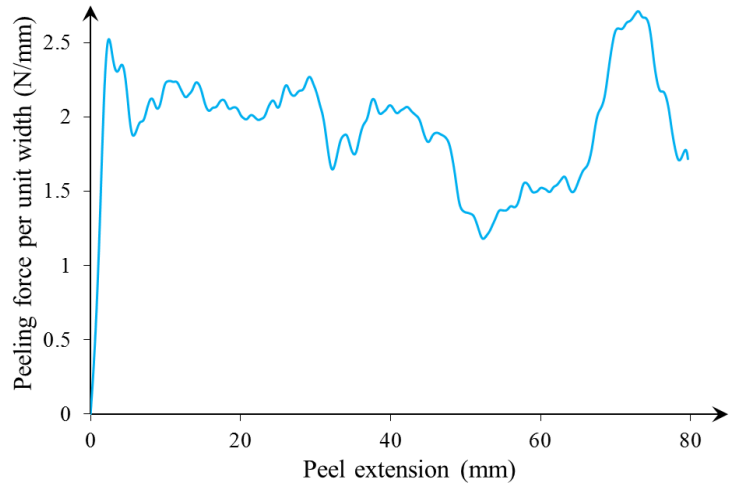

(a)

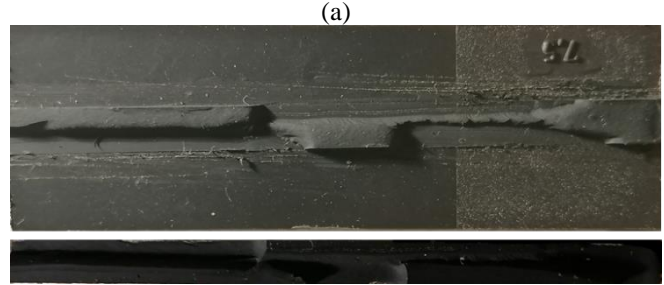

(b)

*Fig. 11. Surface roughness effect on (a): the 90° peeling response, and (b): the failure pattern of specimen with adhesive thickness t=1.9 mm.*

## 4. Conclusion

In the present contribution, the fracture behavior of adhesive joints involving the surface roughness effect has been addressed. To this aim, the peeling response of silicone-based adhesives on smooth and rough substrates has been characterized through 90° and 180° peeling tests. Then, the post-mortem analysis of peeled off specimens has been carried out to lay out the failure modes. Adhesive failure (AF), cohesive failure (CF), and combination of these two modes of failure have been encountered. It is found that, in the 90º peel angle configuration, once damage initiates in a mixed failure mode making a ridge pattern on the substrate, the load to make the crack propagate is higher as compared to the failure mode governed by CF mode. While in the 180° case, there is almost a pure AF or CF close to the strip at the beginning and the required load to make crack propagate in case of CF is much higher (almost two times more) compared to the AF case. However, for the 180° peeling test configuration, regardless of the failure mode at the beginning, the same plateau of peeling force level has been recorded, which represents the same adhesion energy for each similar case. In addition, results reveal that the surface roughness affects the peeling response, especially in the 90º configuration, while it has a negligible role for 180° peeling tests. The reason can be attributed to the fact that crack propagates in the opening mode in the 90º peeling, while mixed-mode failure with shear dominated fracture occurs in the 180º peeling test.

# References

[1]  Borri C, Paggi M, Reinoso J, Borodich FM. Adhesive behaviour of bonded paper layers: Mechanical testing and statistical modelling. Proc Inst Mech Eng Part C J Mech Eng Sci 2016;230:1440–8. https://doi.org/10.1177/0954406215612502.

[2]  Corrado M, Paggi M. Nonlinear fracture dynamics of laminates with finite thickness adhesives. Mech Mater 2015;80:183–92. https://doi.org/10.1016/j.mechmat.2014.07.012.

[3]  Rahmani A, Choupani N. Experimental and numerical analysis of fracture parameters of adhesively bonded joints at low temperatures. Eng Fract Mech 2019;207:222–36. https://doi.org/10.1016/j.engfracmech.2018.12.031.

[4]  Sun S, Li M, Liu A. A review on mechanical properties of pressure sensitive adhesives. Int J Adhes Adhes 2013;41:98–106. https://doi.org/10.1016/j.ijadhadh.2012.10.011.

[5]  Zarei H, Brugo T, Belcari J, Bisadi H, Minak G, Zucchelli A. Low velocity impact damage assessment of GLARE fiber-metal laminates interleaved by Nylon 6,6 nanofiber mats. Compos Struct 2017;167:123–31. https://doi.org/10.1016/j.compstruct.2017.01.079.

[6]  Nimje S V., Panigrahi SK. Interfacial failure analysis of functionally graded adhesively bonded double supported tee joint of laminated FRP composite plates. Int J Adhes Adhes 2015;58:70–9. https://doi.org/10.1016/j.ijadhadh.2015.01.002.

[7]  Kawashita LF, Moore DR, Williams JG. Protocols for the measurement of adhesive fracture toughness by peel tests. J Adhes 2006;82:973–95. https://doi.org/10.1080/00218460600876142.

[8] Wei Y, Hutchinson JW. Interface strength, work of adhesion and plasticity in the peel test. Int J Fract 1998;93:315–33. https://doi.org/10.1007/978-94-017-2854-6_16.

[9] Kim J, Kim KS, Kim YH. Mechanical effects in peel adhesion test. J Adhes Sci Technol 1989;3:175–87. https://doi.org/10.1163/156856189X00146.

[10] Peng Z, Wang C, Chen L, Chen S. Peeling behavior of a viscoelastic thin-film on a rigid substrate. Int J Solids Struct 2014;51:4596–603. https://doi.org/10.1016/j.ijsolstr.2014.10.011.

[11] Peng ZL, Chen SH. Effects of surface roughness and film thickness on the adhesion of a bioinspired nanofilm. Phys Rev E - Stat Nonlinear, Soft Matter Phys 2011;83:1–8. https://doi.org/10.1103/PhysRevE.83.051915.

[12] Huber G, Gorb SN, Hosoda N, Spolenak R, Arzt E. Influence of surface roughness on gecko adhesion. Acta Biomater 2007;3:607–10. https://doi.org/10.1016/j.actbio.2007.01.007.

[13] Patil OR, Ameli A, Datla N V. Predicting environmental degradation of adhesive joints using a cohesive zone finite element model based on accelerated fracture tests. Int J Adhes Adhes 2017;76:54–60. https://doi.org/10.1016/j.ijadhadh.2017.02.007.

[14] Datla N V., Papini M, Ulicny J, Carlson B, Spelt JK. The effects of test temperature and humidity on the mixed-mode fatigue behavior of a toughened adhesive aluminum joint. Eng Fract Mech 2011;78:1125–39. https://doi.org/10.1016/j.engfracmech.2011.01.028.

[15] Datla N V., Ulicny J, Carlson B, Papini M, Spelt JK. Mixed-mode fatigue behavior of degraded toughened epoxy adhesive joints. Int J Adhes Adhes 2011;31:88–96. https://doi.org/10.1016/j.ijadhadh.2010.11.007.

[16] Pickering JP, Van Der Meer DW, Vancso GJ. Effects of contact time, humidity, and surface roughness on the adhesion hysteresis of polydimethylsiloxane. J Adhes Sci Technol 2001;15:1429–41. https://doi.org/10.1163/156856101753213286.

[17] Paggi M, Reinoso J. A variational approach with embedded roughness for adhesive contact problems. Mech Adv Mater Struct 2018;0:1–17. https://doi.org/10.1080/15376494.2018.1525454.

[18] Cho TM, Choo YS, Lee MJ, Oh HC, Lee BC, Park TH, et al. Effect of surface roughness on the adhesive strength of the heat-resistant adhesive RTV88. J Adhes Sci Technol 2009;23:1875–82. https://doi.org/10.1163/016942409X12508517390671.

[19] Zarei H, Marulli MR, Paggi M, Pietrogrande R, Üffing C, Weißgraeber P. Adherend surface roughness effect on the mechanical response of adhesive joints. ArXiv Prepr 2020;arXiv:2006.

[20] Creton C, Ciccotti M. Fracture and adhesion of soft materials: A review. Reports Prog Phys 2016;79:46601. https://doi.org/10.1088/0034-4885/79/4/046601.